# Spatial Quantization: Advancing Insights for Enhancing RRAs Performance

Xiaocun Zong, Fan Yang, *Fellow, IEEE*, Shenheng Xu, *Member, IEEE,* and Maokun Li, *Fellow, IEEE*

*Abstract*—In the new perspective of spatial quantization, this article systematically studies the advantages of reconfigurable reflectarrays (RRAs) designed with closely spaced elements in terms of sidelobe level (SLL) and scanning performance, including theoretical analysis and simulation verification. This article sequentially studies RRAs with element periods of λ/2, λ/4 and λ/8. Both theoretical and simulation results show that under the condition of the same aperture size, with the number of spatial quantization bits increasing, the SLL performance of 1bit RRA using closely spaced structure will have a improvement of about 5.16dB at 0° output direction and 9.34dB at 60° output direction. The scanning accuracy at 60° is improved from 54.52° at λ/2 to 57.97° at λ/8, while the scan loss is reduced from 5.02dB at λ/2 to 3.10dB at λ/8. This study has an important reference value for reconfigurable reflectarray design, communication system and radar design.

## I. INTRODUCTION

RECONFIGURABLE reflectarray (RRA) is a new kind of phased array, thanks to the spatial feeding and integrated phase-tuning technique, RRAs have a simple and low-cost structure compared to conventional phased arrays [1]-[3], and thus have become more and more popular in various civilian and military applications. In the RRA design, the key to beamforming is to assign the required phase value to each element. The most popular method to realize phase distribution is using PIN diodes [4], [5], according to the number of phase states, it can be divided into 1bit, 2bit and even 3bit, then we call this phase quatization.

Inspired by the concept of phase quantization [6], this paper introduces the concept of spatial quantization to describe the varying precision levels of antenna arrays due to differences in element sizes. Traditionally, the period of a reflectarray element is designed to be half the wavelength (λ/2) at the center frequency. However, with advancements in closely spaced element design, antenna arrays with element periods smaller than half a wavelength—referred to as sub-wavelength designs—have been developed [7]-[13].

To quantify spatial precision, we define spatial quantization bits based on the size of the sub-wavelength element. Specifically, a λ/2 element period corresponds to 1-bit spatial quantization. Reducing the element period to λ/4, which is half the size of λ/2, results in 2-bit spatial quantization. Further halving the element period to λ/8 corresponds to 3-bit spatial quantization. In general, each halving of the element period increases the spatial quantization precision by 1 bit.

***Define:*** { N bits spatial quantization: $P_{element} = \frac{\lambda}{2^N}$}

Spatial quantization achieved via closely spaced element design offers significant advantages across various fields. In antenna design, numerous studies [7]-[13] have demonstrated that closely spaced elements can effectively enhance the bandwidth performance of arrays. For instance, in [10], the element period is as small as 0.1λ. Additionally, studies have shown that closely spaced designs contribute to improving the beam scanning range. Reference [14] highlights a sub-wavelength element design that enables 360° beam scanning across the entire space. In the field of communication, closely spaced designs also exhibit notable advantages. References [15]-[17] demonstrate that spatial quantization significantly enhances spectral efficiency, while [15] and [18] further indicate its role in improving channel capacity. Other studies [19], [20] discuss additional applications of closely spaced arrays that are vital for communication systems.

In summary, a systematic study on spatial quantization is essential, as it holds great significance for improving antenna performance, advancing communication systems, and enhancing radar detection capabilities. This article is organized as follows: Section II introduces the basic composition of RRA and the basic working principle of beamforming, as well as some important parameters of RRA; Section III talks about the improvement of SLL performance by spatial quantization; Section IV describes the improvement of beam scanning accuracy and scan loss by spatial quantization; Section V concludes this paper.

## II. BASIC PRINCIPLES OF RRAs

Fig. 1 illustrates the general structure of a reconfigurable reflectarray (RRA) based on a digital phase-controlled electromagnetic surface, consisting of two main components: a feed horn and a digitally controllable electromagnetic surface.

This paragraph of the first footnote will contain the date on which you submitted your paper for review. It will also contain support information, including sponsor and financial support acknowledgment. For example, "This work was supported in part by the U.S. Department of Commerce under Grant BS123456".

The next few paragraphs should contain the authors' current affiliations, including current address and e-mail. For example, F. A. Author is with the National Institute of Standards and Technology, Boulder, CO 80305 USA (e-mail: author@ boulder.nist.gov).

S. B. Author, Jr., was with Rice University, Houston, TX 77005 USA. He is now with the Department of Physics, Colorado State University, Fort Collins, CO 80523 USA (e-mail: author@lamar.colostate.edu).

T. C. Author is with the Electrical Engineering Department, University of Colorado, Boulder, CO 80309 USA, on leave from the National Research Institute for Metals, Tsukuba, Japan (e-mail: author@nrim.go.jp).

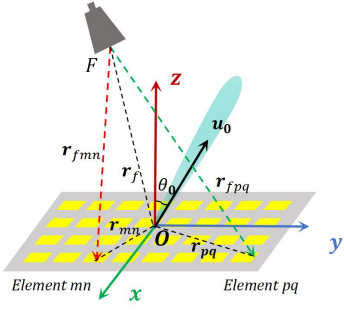

Fig.1. The general structure of reconfigurable reflectarray antenna.

To achieve the focusing of the beam in the $u_0$ direction, the phase of each element on the array surface should form an equal phase surface in the direction perpendicular to $u_0$ after compensating the radiation of the feed source. Assuming that all elements on the surface are located in the far field of the feed source, and each element should provide a phase:

$$\varphi_{mn}^{req} = k \cdot (r_{fmn} - u_0 \cdot \vec{r}_{mn}) + \varphi_0 \quad (1)$$

where $k$ is free-space wavenumber, $r_{fmn}$ is the spatial distance between feed and the mnth element, $r_{mn}$ is the position vector of the mnth element, $u_0$ is the unit vector in the beam direction, $\varphi_0$ is a reference phase constant for all elements. When a specific phase value $\varphi_{mn}^{req}$ is assigned to an element, it is rounded to the nearest quantized phase based on the desired phase requirement. Then the required beam is formed.

According to the array principle, the radiation pattern of the antenna shown in Fig.1 can be expressed as formular (2). With the expression of the pattern, the antenna gain and other parameters can also be calculated.

$$E(\theta, \varphi) = cos^{q_e} \sum_{m=0}^{M-1} \sum_{n=0}^{N-1} A_{mn} e^{-jkr_{fmn}} e^{j\varphi_{mn}} e^{jk\vec{r}_{mn} \cdot u} \quad (2)$$

The spatial quantization concept we proposed further quantizes the $r_{mn}$ parameter. Following the calculation process outlined above, we compute the beam scanning results for different spatial quantization bit resolutions, as illustrated in Fig.2. The results clearly demonstrate that spatial quantization significantly improves sidelobe level, scanning accuracy, and scanning loss.

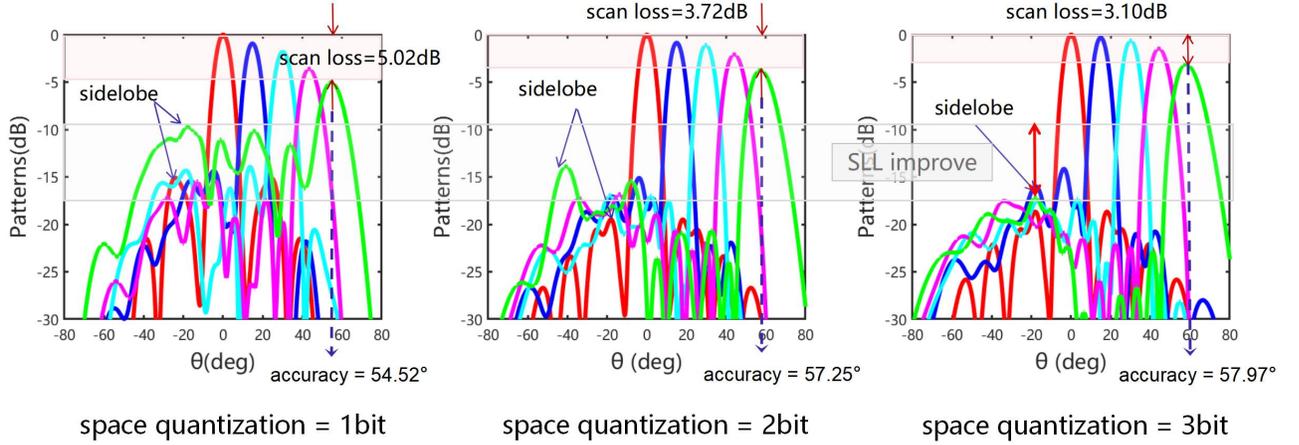

Fig.2. Scanning radiation patterns from 0° to 60° with a 15° step under different spatial quantization bit.

### III. IMPROVEMENT OF SLL PERFORMANCE BY SPATIAL QUANTIZATION

Sidelobe level is an important parameter for judging RRA performance. We define the SLL to be the maximum value of the total radiation pattern out side of the main beam [21]. In this section, we first study the SLL improvement of 1-bit RRA with spatial quantization bits increasing; then we give the reasons for the improvement of SLL performance via spatial quantization.

A 1-bit RRA is simulated using the array method, ensuring that the aperture size remains constant for arrays with different element periods. For an element period of λ/2, the array size is 16×16; for λ/4, the array size is 32×32; and for λ/8, the array size is 64×64.

As the spatial quantization accuracy improves, the SLL performance gradually enhances, as shown in the pattern results in Fig.3 and TableI. When the outgoing beam is at 0°, the SLL for λ/4 is 4.46dB better than λ/2, and λ/8 is 5.16dB better than λ/2. When the outgoing beam is at 30°, λ/4 performs 3.71dB better than λ/2, and λ/8 is 4.53dB better. At 45°, λ/4 improves by 3.03dB over λ/2, and λ/8 performs 4.35dB better. Finally, when the outgoing beam is at 60°, the SLL of λ/2 is -4.69dB, while λ/4 is 5.47dB better and λ/8 is 9.34dB better than λ/2.

TABLE I
SIDELOBE LEVELS CONTRAST AT DIFFERENT BEAM ANGLES WITH DIFFERENT SPATIAL QUANTIZATION FOR 1BIT RRAs

| Scan Angle | SLL with differnt spatial quantization / better than λ/2 (dB) | | |
|---|---|---|---|
| | λ/2 | λ/4 | λ/8 |
| 0° | -14.00 | -18.46/4.46 | -19.16/5.16 |
| 30° | -12.15 | -15.86/3.71 | -16.68/4.53 |
| 45° | -11.72 | -14.75/3.03 | -16.07/4.35 |
| 60° | -4.69 | -10.16/5.47 | -14.03/9.34 |

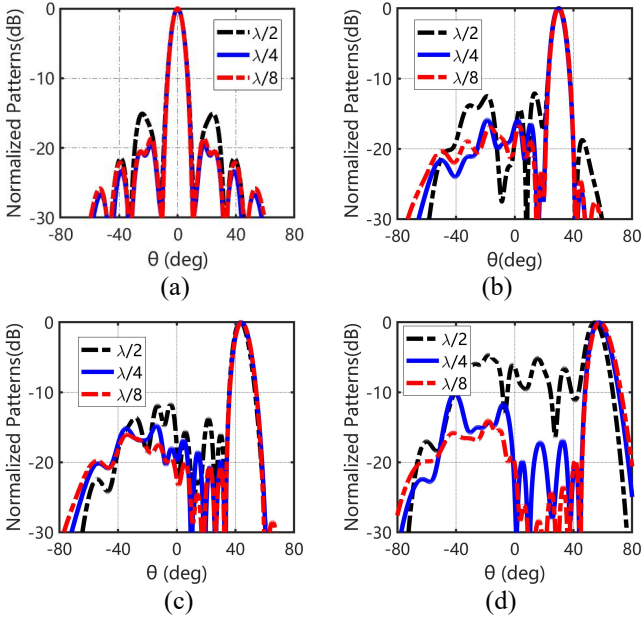

Fig.3. Normalized radiation patterns for RRAs at different spatial quantization levels: (a) point at 0°; (b) point at 30°; (c) point at 45°; and (d) point at 60°.

It can be observed that as the emission angle increases, the SLL performance deteriorates, regardless of the spatial quantization precision. Spatial quantization significantly enhances SLL performance. When the emission angle is small, such as at 0°, the SLL improvements for λ/4 and λ/8 are similar. However, as the emission angle increases, the higher the spatial quantization accuracy, the better the SLL sidelobe performance. In summary, spatial quantization can improve SLL performance by 3 to 10dB, depending on the scan angle and the number of spatial quantization bits.

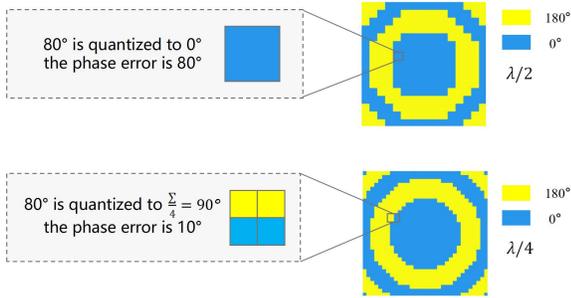

Fig.4. The phase error caused by spatial quantization is reduced.

The reason why SLL performance is getting better with the improvement of spatial quantization accuracy is that the finer the array division, the smaller the phase error will be at the phase boundary, and the phase will be closer to the theoretical value, so the calculated directional pattern will be closer to the ideal situation. For example, as shown in Fig.4, the theoretical calculated phase at a certain element of the λ/2 array is 80°. At this time, the phase is 0° after 1-bit quantization, and the error caused by phase quantization is 80°; but if the array scale is λ/4, one element is divided into four elements, and the element phase layout is more precise. After 1-bit quantization, the phase of two elements is 0° and the phase of two elements is 180°, the average phase is 90°, and the error is 10°. Therefore, spatial quantization will make the phase calculation at the phase boundary more accurate, thereby making the SLL performance better. When the emission angle increases, there will be more phase boundaries, so the phase accuracy will be further improved and the effect of spatial quantization will be more significant.

A theoretical explanation of the phenomenon where spatial quantization reduces the SLL is provided, using the normal outgoing beam as a reference. Based on the phase distribution of the elements, we partition the phase region into two distinct areas: S2 is the area where the phase remains unaffected by spatial quantization, and S1 is the boundary region where phase changes occur due to spatial quantization. The impact of these two regions on the array radiation pattern is then analyzed. According to the two regions into which the array is divided, the array pattern calculation can also be written accordingly:

$$E(\theta,\varphi) = cos^{q_e}\theta \left( \sum_{\substack{m,n \in S_1 \\ 1 \leq m,n \leq M,N}} + \sum_{\substack{m,n \in S_2 \\ 1 \leq m,n \leq M,N}} \right) \quad (3)$$

As shown in Fig.5, after spatial quantization, the patterns of elements whose phases remain unchanged are unaffected by the quantization process. However, for the elements located near the phase boundary, spatial quantization causes the phase distribution to better approximate the theoretical values, resulting in pattern with lower side lobe. When the two patterns are superimposed, the improvement in sidelobe performance due to spatial quantization becomes evident.

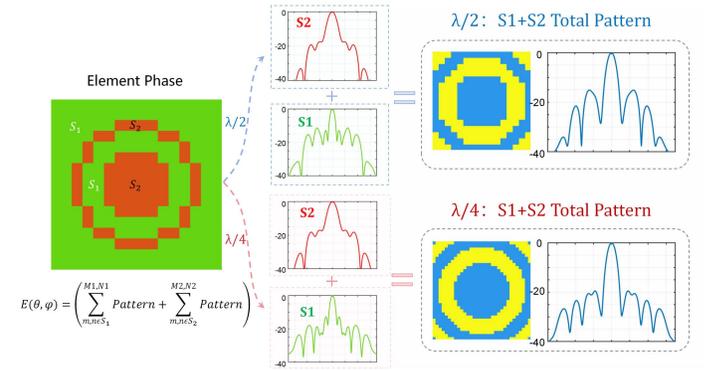

Fig.5. The process of synthesizing the pattern from two regions and the comparison of different spatial quantization.

## IV. IMPROVEMENT OF SCANNING ACCURACY AND LOSS BY SPATIAL QUANTIZATION

Scanning accuracy and scan loss are also important criteria for measuring RRA performance, and have very high index requirements in applications such as radar detection. In order to explore the impact of spatial quantization on scan accuracy and scan loss, this section makes a detailed analysis.

As the spatial quantization accuracy increases, the element size continues to decrease, the directivity weakens, and the equivalent qe value also decreases. We choose *TLX-8* as the dielectric material, and used *HFSS* to simulate the radiation pattern of λ/2 period , λ/4 period and λ/8 period, the $q_e$ values are 2.5, 1.5 and 1 respectively, which is shown in Fig.6. To ensure the rationality of subsequent simulations, we set the calculated parameter $q_e$ to be consistent with the $q_e$ value obtained by *HFSS* simulation, this is used to distinguish the different directivity between elements, improving the consistency between the simulation and the actual situation.

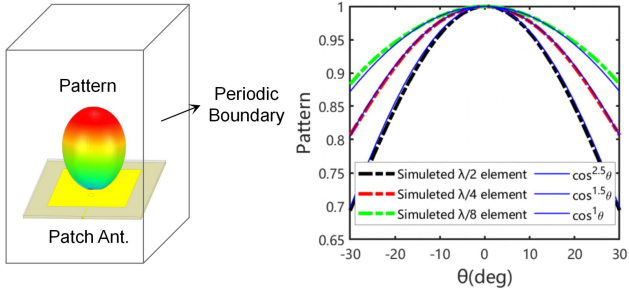

Fig.6. Element pattern $q_e$ values simulated in *HFSS*.

*A. Scanning Accuracy*

The theoretical calculated phase result of 60° beam emission is used when designing the phase layout. When the scanning angle reaches 60°, the scanning accuracy continues to improve as the number of phase quantization bits increases, it can be seen from Fig.7. The scanning accuracy at 60° is improved from 54.52° at λ/2 to 57.97° at λ/8. The larger the scanning angle, the more obvious the advantage of spatial quantization. Of course, the error in scanning accuracy can be compensated by optimizing the phase. If we arrange the phase according to 64°, it will scan to 60° probably, thus compensating for the defect in scanning accuracy.

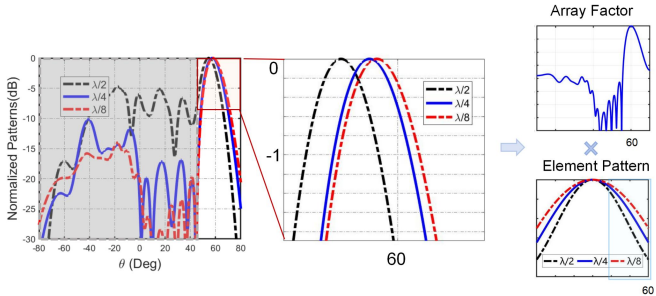

Fig.7. Scanning error under different spatial quantization accuracy.

This phenomenon occurs because, at a spatial quantization of λ/2, the element exhibits strong directivity. Consequently, the element's inherent radiation level is significantly reduced at large angles. As a result, the array's performance in large-angle scanning deteriorates, leading to reduced scanning accuracy. Additionally, this causes a noticeable drop in gain at large angles and a higher scan loss.

*B. Scanning Loss*

Spatial quantization can also improve the performance of scan loss. When we usually do RRA testing of half-wavelength design, we often encounter that the gain of large-angle scanning is much lower than the gain of normal beam. Simulation results show that spatial quantization can improve this problem well. It can be seen from Fig.8 and Table II, increasing the number of spatial quantization bits results in a reduction of the element period from λ/2 to λ/8. This leads to a slower variation in the element gain with respect to the angle. Consequently, the scan loss during scanning is minimized. Furthermore, the figures clearly demonstrate that spatial quantization enhances the sidelobe level performance.

TABLE II
GAIN AND SCAN LOSS AT DIFFERENT BEAM ANGLES WITH DIFFERENT SPATIAL QUANTIZATION FOR 1BIT RRAS

| Period | Gain and scan loss (dB) | | | | |
|---|---|---|---|---|---|
| | 0° | 15° | 30° | 45° | 60° |
| λ/2 | 25.91/- | 24.98/0.93 | 24.10/1.81 | 22.32/3.59 | 20.89/5.02 |
| λ/4 | 25.87/- | 25.02/0.85 | 24.80/1.07 | 23.80/2.07 | 22.15/3.72 |
| λ/8 | 25.64/- | 25.32/0.32 | 25.03/0.61 | 24.18/1.46 | 22.54/3.10 |

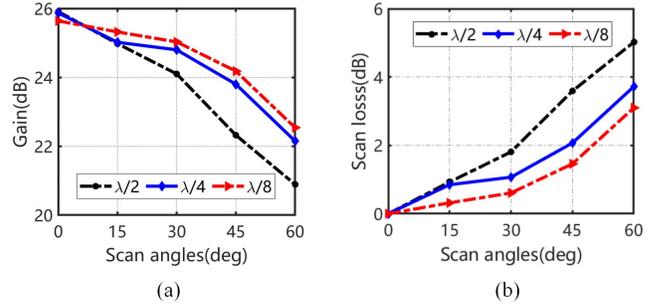

Fig.8. Gain and scan loss from 0° to 60° with a 15° step under different spatial quantization accuracy: (a) gain, (b) scan loss.

As illustrated in Fig.8, when the array scanning angle for an element period of λ/2 changes from 15° to 60°, the scan loss increases from 0.93 dB to 5.02 dB. For an element period of λ/4, the scan loss rises from 0.85 dB to 3.72 dB over the same range, and for an element period of λ/8, the scan loss increases from 0.32 dB to 3.10 dB. These results indicate that a higher number of spatial quantization bits results in smaller variations in array gain with respect to the scanning angle, thereby reducing scan loss.

V. CONCLUSION

This article provides a comprehensive analysis of the impact of spatial quantization on the performance of reconfigurable reflectarrays. While previous studies have primarily focused on bandwidth performance via closely spaced element design, this work highlights the significant improvements that spatial quantization can achieve in reducing sidelobe levels. Furthermore, spatial quantization demonstrates clear advantages in enhancing beam-scanning capabilities, including improved scanning accuracy and reduced scan loss. These findings offer valuable insights for the development of communication systems and radar detection applications.